# Bioabsorbable WE43 Mg alloy wires modified by continuous plasma-electrolytic oxidation for implant applications. Part I: processing, microstructure and mechanical properties

Wahaaj Ali[1,2,3], Muzi Li[1], Leon Tillmann[3], Tim Mayer[3], Carlos González[1,4], Javier LLorca[1,4*] and Alexander Kopp[3]

[1]IMDEA Materials, C/Eric Kandel 2, 28906 Getafe, Madrid, Spain
[2]Departament of Material Science and Engineering, Universidad Carlos III de Madrid, Leganés, Madrid 28911, Spain
[3]Meotec GmbH, Philipsstr. 8, 52068 Aachen, Germany
[4]Department of Materials Science, Polytechnic University of Madrid/Universidad Politécnica de Madrid, 28040 Madrid, Spain

## Abstract

In our work, a novel processing strategy for the continuous fabrication and surface modification of wires from Magnesium alloy WE43 by means of plasma-electrolytic oxidation (PEO) is presented. In the first step, wires with a strong basal texture and small grain size (≈ 1 µm) were manufactured by combined cold drawing and in-line stress-relief heat treatment steps that optimized the mechanical properties (in terms of strength and ductility) by means of annealing. In a second step, and to the best of our knowledge for the first time ever, the wires were continuously surface-modified with a novel plasma electrolytic oxidation process, which was able to create a homogeneous porous oxide layer made of MgO and $Mg_3(PO_4)_2$ on the wire surface. While the oxide layer slightly diminished the tensile properties, the strength of the surface-modified wires could be maintained close to 300 MPa with a strain-to-failure ≈ 8%. Furthermore, the thickness of the oxide layer could be controlled by immersion time within the electrolytic bath and was adjusted to realize a thicknesses of ≈ 8 µm, which could be obtained in < 20 s. Our experiments showed that the chemical composition, morphology and porosity of the oxide layer could be tailored by changing electrical parameters. The combined cold drawing and heat treatment process with additional continuous plasma electrolytic oxidation processing can be upscaled to produce a novel generation of bioabsorbable Mg wires with optimized mechanical, degradation and biological performance for use in biomedical applications.

1. **Introduction**

Bioresorbable metallic wires of iron (Fe) [1], zinc (Zn) [2], and magnesium (Mg) [3] have been investigated for a broad range of biomedical applications. In particular, Mg wires have received the most attention due to their suitable mechanical properties (elastic modulus of 40-45 GPa and tensile strength in the range 100-600 MPa) [4], faster degradation rate than their counterparts (iron, zinc, and molybdenum) and biocompatibility [5]. The initial applications envisaged for Mg wires include staples, sutures and meshes for wound closure [6], cardiovascular stents [3], bioresorbable electronic sensors [7], tumor treatment [8], etc. Mg fibers (or wires) of 0.10 mm [9] and 0.30 mm [10] diameter have been also used as unidirectional reinforcement of poly-lactic acid polymers [11]), and excellent for manufacturing bioresorbable composites with improved mechanical properties and degradation performance for orthopedic applications. In addition, more exotic biomedical functions of Mg wires have been recently explored. For instance, Qiao et al. [12] reported the antitumor properties of Mg wires of 0.80 mm in diameter in a subcutaneous tumor model as Mg degradation by-products inhibited the growth of SKOV3 cells. Hopkins et al. [13] further successfully placed short Mg filaments (made from 0.25 mm of diameter wires) inside a hollow nerve conduit to support long and short gap nerve repair. Filaments were partially absorbed and showed good attachment of regenerating cells for short gaps after 6 weeks. Most recent Xue et al, proposed 3D-weaved scaffolds (made from 0.25 mm and 0.30 mm diameter wires) from Mg ZXM100 wires with controlled porosity as an alternative to complex additive manufacturing routes [14].

Manufacturing of wires from pure Mg [15] as well as Mg alloys (ZEK100, MgCa0.8, AL36 [6] and AZ31, AZ80, AZ91 [16]) has been reported by direct extrusion at high temperature (230ºC-450ºC). However, this process limits the minimum diameter of the wire (often > 500 µm) because is unstable for thin wires and not to be very robust in terms of geometrical accuracy. Cold drawing still depicts the standard manufacturing method to process metallic wires [17] although the limited reduction of the diameter during cold drawing requires many passes to obtain very thin wires and -as a result- long manufacturing times. The main advantage of cold drawing -with respect to hot drawing or hot extrusion- is the higher strength of the wires (because of the smaller grain size) and the possibility to reach very small diameters (< 300 µm), which is critical for many applications that require flexible wires (including additive manufacturing by fused filament fabrication). Moreover, cold drawing is limited by the strong plastic anisotropy of Mg alloys, which is associated with their hexagonal close packed lattice structure [18]. The critical resolved shear stress to active basal slip is much lower than necessary to promote prismatic or pyramidal slip or tensile twinning [19][20]. As a result, cold drawing of Mg leads to the development of a strong texture in which the basal planes are parallel to the drawing axis and further plastic deformation requires large stresses to activate the hard slip systems, leading to large stress concentrations and often to fracture. This plastic anisotropy can be reduced by the addition of different alloying elements [19]. Cold drawing of thin wires (< 0.50 mm) of different Mg alloys has been reported in the literature, including Mg-Zn-Ca [21] and AZ31 [20]. Moreover, the microstructure of the Mg wires processed by cold drawing can be tailored through thermal treatments [4][22][23][24] to optimize mechanical properties and their corrosion rate.

Cold drawing of Mg alloys containing rare earths (Mg-RE) is favored by the reduction in texture [25] and shows potential to realize thin wires, which have recently been reported by several groups [4] [6] [26] [27][28]. The presence of RE improves the strength of Mg alloys [29] as well as the corrosion resistance [30]. Moreover, WE43 and other RE containing Mg alloys have used within approved cardiovascular and orthopedic devices [31]. It seems well established that annealing of Mg-RE wires between 250ºC and 450ºC can be used to tailor the mechanical properties and corrosion rates of these materials [4] [22]. However, most of the literature does not provide enough detailed description of cold drawing parameters to reproduce thin wires made of Mg-RE alloys. In addition, the influence of cold work and of annealing time and temperature on dislocation structures, texture and spatial distribution of precipitates has not been analyzed to such an extent as to properly understand how processing properties link to strength and corrosion in these materials.

Despite the increasing number of potential applications, the current limitations of Mg wires as bioabsorbable materials need to be critically evaluated and taken into consideration. The first – and the most relevant one – is the high corrosive reactivity of Mg within the body, which leads to an increase in the local pH due to a higher concentration of $OH^-$ ions and to the formation of $H_2$ gas accumulations that – above certain thresholds – can cause an inflammatory response of the surrounding tissue [22]. This problem is particularly relevant in the case of Mg wires or porous scaffolds with a high specific surface, in which fast degradation rates might not allow for proper cell attachment that can further act as corrosion protection and slow down degradation [31]. In addition, magnesium dissolution as any other metallic alloy is generally associated with pitting corrosion, as reported in numerous *in vitro* studies [9], which in particular affects the mechanical integrity of thin wires and makes it difficult to ensure that the reduction in strength over time is matched with device requirements and ingrowth of surrounding tissue. Improvements in corrosion resistance and in severity of pitting corrosion of Mg wires have been reported through alloying with RE and application of suitable thermal treatments [4][22][23][24]. Much larger reductions in reactivity seem, however, necessary for the development of the next generation Mg wire-based implants that meet all requirements for potential biomedical applications.

In this context, surface modification is known to be the most effective strategy to control Mg corrosion, i.e. degradation [32]. Among different possibilities, Plasma Electrolytic Oxidation (PEO) showed most suitable to significantly reduce the degradation rate while concurrently improving the cytocompatibility of magnesium implants [5] [33]. A micro-porous oxide layer is formed on the surface during PEO which serves as an effective barrier between the Mg substrate and the surrounding physiological environment. The morphology of this oxide layer can be controlled by various variables, e.g. electrical parameters (voltage, current, frequency, etc.) or electrolyte composition, which both affect the barrier properties as well as the overall cytocompatibility of such surface-modified Mg alloys [34]. While PEO, also known as Micro-Arc Oxidation (MAO), for Mg alloys is generally well established [35][36], Chu et al. [37] were the first to apply this technique to AZ31 Mg alloys wires. They reported that wires modified by PEO offered better protection against corrosion and retarded surface degradation in different simulated physiological environments (simulated body fluid and simulated intestinal fluid) at pH of 3 and 7. Nevertheless, they were limited by conventional PEO processing that only allows for the modification of distinct segments of wire [37] that fit into the electrolyte bath, but cannot be scaled up to larger amounts by its discontinuous nature of processing.

To overcome these limitations, our work aimed to develop a continuous processing route combining cold drawing, thermal annealing and surface-modification by PEO to optimize the microstructure and in particular corrosion properties of WE43 Mg wires as well as to provide sufficient amounts of surface modified wire for the textile processing of potential biomedical devices, e.g., by braiding. In this first part of our comprehensive work, the cold drawing/annealing process is described in detail, comprising a new and innovative manufacturing strategy for continuous surface-modification of Mg wires by PEO. This new strategy seems suitable for upscaling and thus industrial application using large quantities of Mg wires and seems to be the first to provide high quality surface modification of metallic wires by PEO using a continuous process. While the relationship between the processing conditions (degree of cold work, annealing temperature and time, PEO parameters) and the microstructure of the wires (grain size, dislocation and precipitate structures, oxide layer thickness, porosity and composition) as well as tensile behavior is reported in this first part of our work, further links between the microstructure and the degradation and biological performance of the wires is reported in the second part of our work [38].

## 2. Materials and Experimental techniques

### 2.1. Manufacturing of Mg wires by combined cold drawing and heat treatment

In a first step, billets of Mg alloy WE43MEO with nominal composition 1.4 – 4.2 % Y, 2.5 - 3.5 % Nd, <1 % (Al, Fe, Cu, Ni, Mn, Zn, Zr) and balance Mg (in wt.%) estimated by ICP-OES (Varian 720-ES, Varian Inc, USA) were produced by directional gravity chill casting, homogenized by T4 heat treatment and machined to remove surface oxidation and allow for indirect extrusion to rods with a diameter of D = 6 mm (Meotec GmbH, Aachen, Germany). In the second step, these rods were cold drawn to wires of D = 1.5 mm (Fort Wayne Metals Research Products Corp., Indiana, USA) which then were used as precursor material for succeeding steps.

The Mg wires of diameter D = 1.50 mm were then cold drawn using an oil-based lubricant (Drawlub AL 4686, PETRFER, Chemie H. R Fischer GmbH + Co. KG, Germany) in 28 steps to achieve the final diameter D= 0.30 mm following the drawing schedule in Fig. 1. Schematic depiction of the cold drawing process. The wire having diameter $D_1$ passes through the die by applying force F to reach the final diameter $D_2$. The half angle used for the die was α= 6º.

Table 1. The die was made of carbide material with 6° half angle (Fig. 1) [39]. The drawing speed for the first pass was 0.8 m/min and 4 m/min for the last pass. The speed was increased proportionally according to the ratio between the surface area to volume (1/D) for intermediate passes. A brief annealing of the wire at 450°C was performed after every pass within an in-line positioned throughput tubular oven of 350 mm in length at the same speed as of the preceding drawing pass, leading to a continuous heat treatment process and thus decreasing necessary off times for annealing the coiled wires by heat treatment after every pass. Finally, lubricant traces were removed by passing the wires continuously through an ultrasonic bath filled with ethanol.

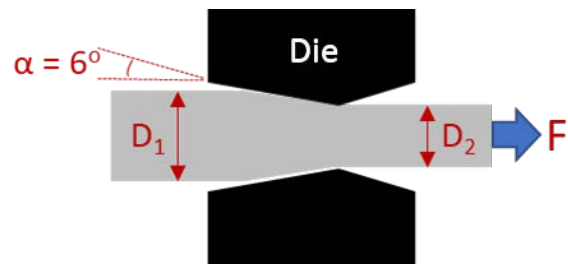

**Fig. 1**. Schematic depiction of the cold drawing process. The wire having diameter $D_1$ passes through the die by applying force F to reach the final diameter $D_2$. The half angle used for the die was α= 6°.

**Table 1.** Drawing scheme for the manufacturing of Mg wires

| Pass | 1 | 2 | 3 | 4 | 5 | 6 | 7 | 8 | 9 | 10 |
|---|---|---|---|---|---|---|---|---|---|---|

| Pass  |       |       |       |       |       |       |       |       |       |
|-------|-------|-------|-------|-------|-------|-------|-------|-------|-------|
| D (mm) | 1.483 | 1.414 | 1.348 | 1.285 | 1.225 | 1.168 | 1.114 | 1.062 | 1.013 | 0.966 |
| Pass  | 11    | 12    | 13    | 14    | 15    | 26    | 17    | 18    | 19    | 20    |
| D (mm) | 0.921 | 0.878 | 0.837 | 0.798 | 0.744 | 0.694 | 0.647 | 0.603 | 0.563 | 0.525 |
| Pass  | 21    | 22    | 23    | 24    | 25    | 26    | 27    | 28    |       |       |
| D (mm) | 0.489 | 0.456 | 0.425 | 0.397 | 0.370 | 0.345 | 0.322 | 0.300 |       |       |

Mg wires of 0.30 mm in diameter were successfully manufactured to an extent of 34% and 13% cold work (CW34 and CW13). In case of CW34, the last three passes were achieved without intermediate annealing and, thus, more cold work was introduced to wires. With regards to length, Mg WE43MEO alloy wires of 0.30 mm in diameter were successfully manufactured up to lengths of 20 m. However, it should be noted that reduction of wire breaks could be significantly reduced, and longer lengths of Mg wire achieved by using diamond dies, optimization of annealing conditions, and a more suitable lubricant in future endeavors. The CW13 wires were later subjected to annealing heat treatment at 400ºC for 5 s or 10 s (HT400-5 and HT400-10) or at 450ºC for the same durations (HT450-5 and HT450-10). The brief annealing times of 5 s and 10 s were chosen to match the facilitated drawing speeds.

## 2.2. Surface-modification by PEO

The demand for surface-modified magnesium wire is expected to significantly increase in future. Therefore, a scalable process for continuous production of PEO surface-modified Mg wires is an important milestone to pave the way for the industrial application of PEO-modified bioabsorbable wires. Conventionally, Mg wires cannot be surface-modified by PEO in continuous mode as the wire can burn, i.e. deteriorate, in three locations [35]: (1) at the anode/wire interface, (2) at the electrolyte/air interface, and (3) within the electrolyte. The first case terminates the success of processing because the wire might break in two, while the other two cases severely damage the surface modification by black burn marks, but do not intersect the continuously fed wire. To overcome these limitations, the process was modified by (1) shifting anode/wire contact outside the electrolyte bath and (2) optimizing the surface modification speed as well as electrical parameters.

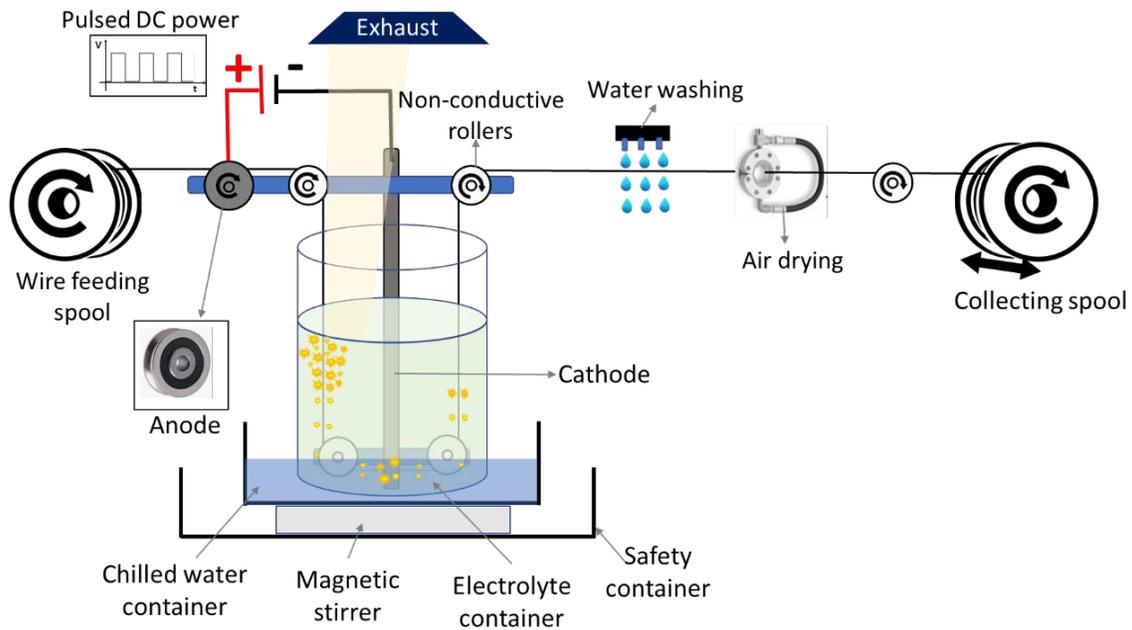

**Fig. 2.** Schematic depiction of the continuous PEO process for surface-modification of Mg wires. Yellow flashes inside the bath show the position and intensity of sparks when the wire passes through the bath in the presence of an electric field.

The processing scheme is depicted in Fig. 2. The spooled HT450-5 Mg wire with 0.30 mm in diameter was mounted on an insulated structure and fed by a stainless-steel roller, which acted as the anode. A vertical stainless steel fixture, that acted as the cathode, was placed inside the phosphate-based electrolyte bath (Kermasorb®, Meotec GmbH, Aachen, Germany). Electrical current was supplied using a power source (LAB/HP15600, ET System Electronic GmbH, Germany) in pulsed mode with a current density of 15A/dm$^2$ at either 250

Hz (PEO250) or 500 Hz (PEO500) to study the effect of frequency on surface quality. The bath temperature was regulated between 20°C to 34°C by suspending chilled water around the bath and constant stirring of the electrolyte with a magnetic stirrer inside the reaction vessel. The surface modification speed was kept at 1.35 m/min leading to the wire spending approximately 20 s within the bath. After leaving the electrolyte, the wire was rinsed by a water jet generated by a pump (MS-4/6 Analog Pump, Ismatec™, Germany) to remove residuals of the electrolyte, dried in air (Air Wipe 009, Airmasters, Germany), and continuously coiled on a spool at the end of the bench. The setup was carefully designed and encapsulated, ensuring electrical safety and the proper collection of process gases by a flexible exhaust system.

## 2.3 Tensile properties

The mechanical properties of wires ($n = 3$) in tension were measured on an electro-mechanical testing machine (ProLine Z010, ZwickRoell GmbH & Co. KG, Ulm, Germany) at room temperature. The gauge length of the wires was 200 mm, and the tests were carried out under displacement control at an average crosshead speed of 5 mm/min with a 10 kN load cell. The mean tensile properties of different wires were compared using one way ANOVA in MS Excel. Data plotting was performed utilizing Python programming.

## 2.4 Microstructural characterization

The grain size and texture of the wires were determined by means of electron backscatter diffraction (EBSD). To this end, wire cross sections of the wires were mounted in a conductive resin, mechanically grounded with SiC paper to 4,000 grit, and then polished with 3 μm and 1 μm diamond paste. The polished samples were etched for 10s with an etchant consisting of 1% $HNO_3$, 24% distilled water and 75% ethylene glycol. The EBSD scan was carried out with an Apreo 2S LoVac FE-SEM (ThermoScientific, Netherlands) equipped with an Oxford-HKL EBSD detector at 20 kV and 2.7 nA with a step size of 0.2 μm. The grain size and inverse pole figures were obtained by post-processing the EBSD map with the software Oxford® HKL Channel-5 (Oxford Instruments Limited, Oxford, United Kingdom).

The dislocation structures and precipitates were analyzed by means of transmission electron microscopy (TEM). Thin foils (< 100 nm) were prepared from the transverse cross-section of the wires by focused ion beam (FIB) in an FEI Helios NanoLab 600i dual-beam FEI-SEM (FEI, Netherlands). TEM analysis was carried out in a FEI Talos (FEI, Netherlands) equipped with a field emission gun operating at 200 kV. Both bright field (BF) images in TEM mode and high-angle annular dark field (HADDF) images in scanning-transmission electron microscopy (STEM) mode were acquired. Chemical composition mapping was performed under STEM mode with a SuperX Energy dispersive X-ray (EDX) detector. The acquisition time was set between 20 and 30 min, and the data were analyzed with the Bruker QUANTAX software.

To analyze oxide layer created by PEO, the wires were mounted in the epoxy resin and grounded to 4,000 SiC paper, followed by polishing with 3 μm diamond suspension and 0.2 μm SiO suspension. They were gently cleaned with ethanol and sputtered with Au. The thickness of the oxide layer created by PEO was estimated from at least 15 measurements randomly taken from the cross-sectional view with an optical microscope (VK-X3000, Keyence Deutschland GmbH, Neu-Isenburg, Germany). The microstructural features of the oxide layer were analyzed in SEM (Zeiss EVO MA15, Germany) at an accelerated voltage of 15~20 kV. Metallograhic sections of the wires in the longitudinal and transverse directions were analyzed in the SEM using secondary electrons as well as energy dispersive X-ray microanalysis (EDX) to ascertain the composition of the oxide layer.

## 3. Results

### 3.1. Effect of processing conditions on the microstructure and mechanical properties of Mg wires

Processing of the Mg wires by cold drawing as described above led to a microstructure with small grain size and a strong basal texture, as it has already been reported in other investigations [4][6][22][23][26]. An example of the microstructure of the wires in the transverse cross-section can be found in the EBSD map in Fig. 3, which corresponds to the wire HT450-5 (13 % cold drawing followed by annealing at 450 ºC during 5 s). The average grain size was 1.4 ± 0.7 µm, and the wires showed a strong texture with the prismatic {10-10} planes parallel to the cross-section and the basal planes parallel to the drawing direction. The EBSD maps of the wires subjected to 34% cold-drawing without additional heat treatment, as well as of wires subjected to 13% cold drawing and additional annealing for different times and temperatures were similar and are not plotted here for the sake of brevity. The grain size of the wire cold drawn up to 34% was slightly smaller (< 1 µm) and presented a similar texture, indicating that annealing did not lead to significant changes in the grain structure. The EBSD maps did not show any evidence of recrystallization during annealing, but large differences in orientation were found inside each grain (denoted by the changes in the color). Those are indicative of strains gradients within the grains, which are associated with large densities of geometrically necessary dislocations near the grain boundaries (Fig. 3a).

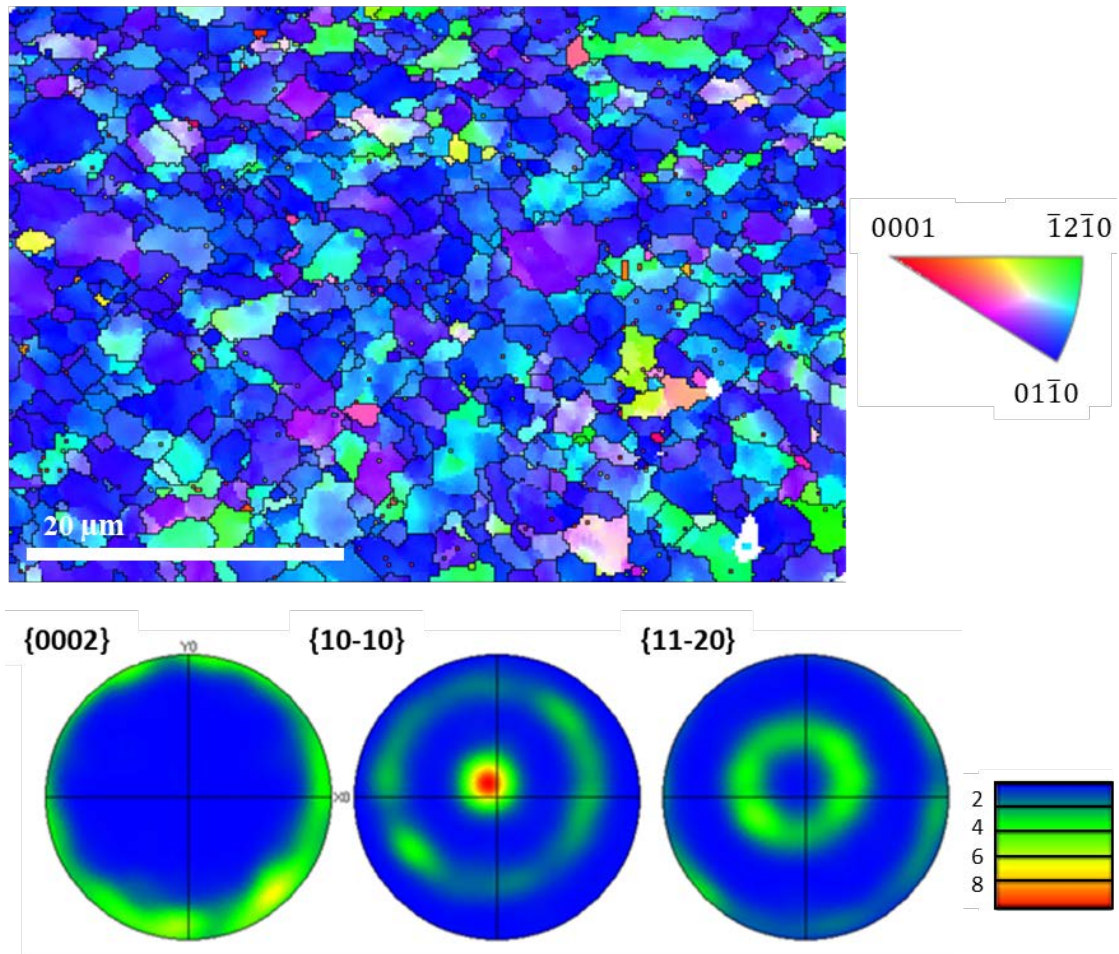

**Fig. 3.** (a) EBSD map of the transverse section of the HT450-5 wire showing the grain structure and orientation with respect to the wire axis. (b) Inverse pole figures showing the orientation of the different planes with respect to the wire axis. The numbers in the legend stand for multiples of random distribution.

The microstructure of the transverse cross-section of the wires was analyzed at higher magnification by means of TEM. Bright-field TEM micrographs of the CW34 and HT450-5 wires are shown in Figs. 4a and 4b, respectively. Large globular shape precipitates (marked with yellow arrows) with an average diameter of 0.5 to 1.5 µm can be found in both microstructures. The EDX analysis of the precipitate composition in Fig. 3b shows that the large precipitates were rich in Nd and Y. Previous investigations of precipitates found in WE43 Mg alloys [31][40] reported that these are β' precipitates with orthorhombic structure and potential composition of Mg$_{12}$YNd. Obviously, the presence of these precipitates was not triggered by the annealing treatments after cold drawing, as they were found in the CW34 wire that was not annealed after cold drawing.

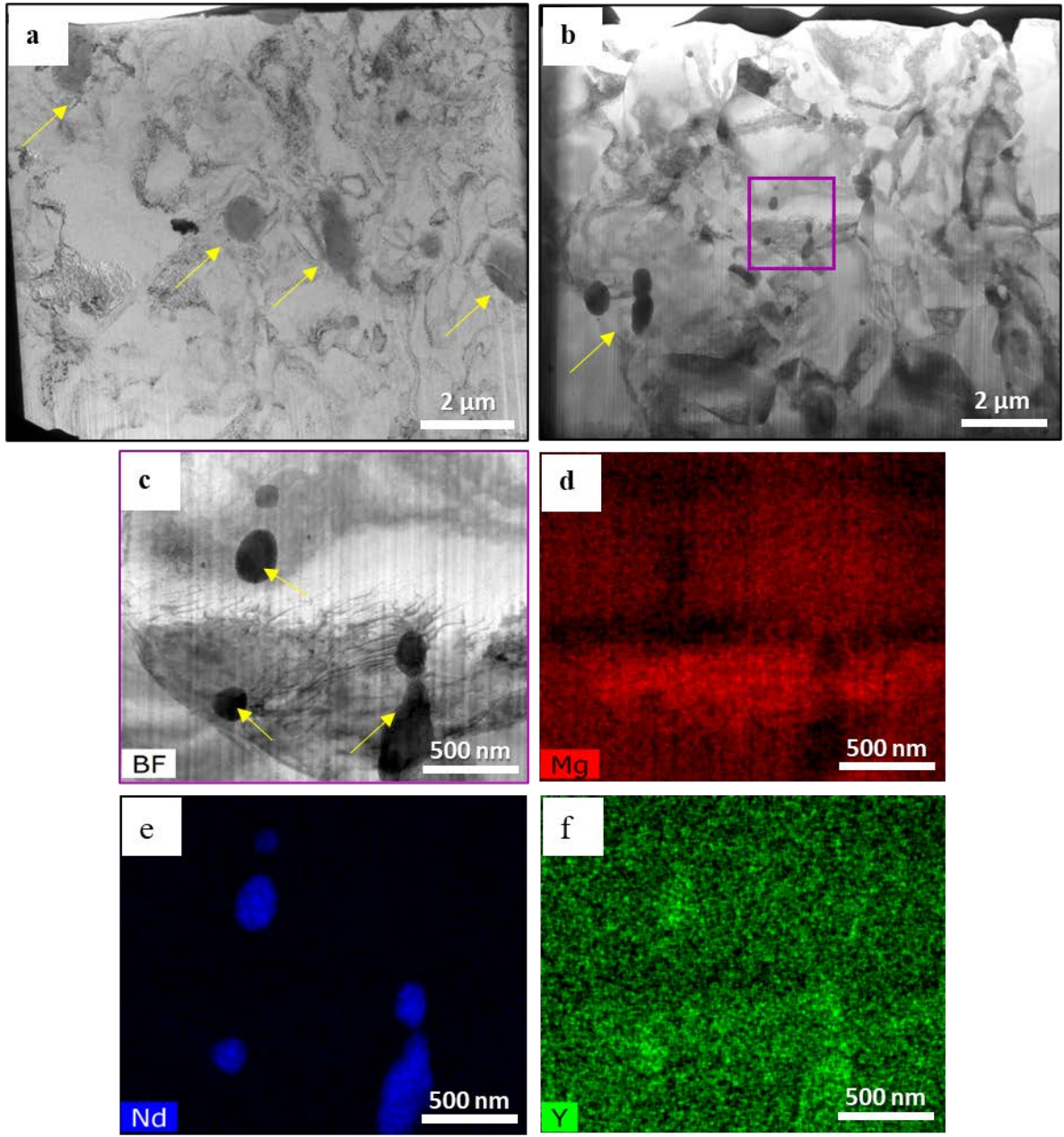

**Fig. 4.** (a) Bright-field TEM image of the transverse section of the CW34 wire. (b) *Idem* of the HT450-5 wire. Large globular precipitates are indicated by yellow arrows in both images. (c) High resolution bright-field TEM of the region marked by a purple square in (b) and (d-f) EDX results show the chemical composition of the large globular precipitates, which are rich in Nd and Y.

Large dislocation densities are found near the grain boundaries in the wire cold drawn up to 34 % without annealing (Fig. 5a). Short annealing of the wires after cold drawing has two main effects. The first one is large reduction in the dislocation densities near the grain boundaries, as shown in Fig. 5b for the HT450-5 wire being annealed for 5s at 450ºC. However, annealing did not remove all the dislocations as temperature and annealing time apparently were not large enough to promote recrystallization of the microstructure, as indicated by the grain structure in Fig. 3a.

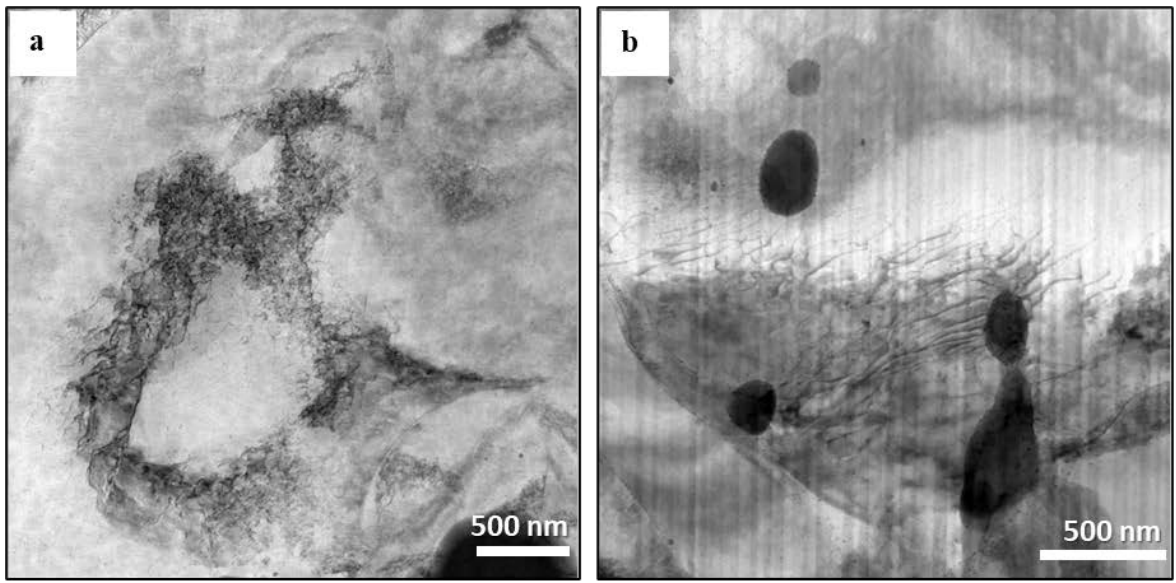

**Fig. 5.** Bright-field TEM images of transverse cross-section of the WE43MEO Mg wires, showing dislocation structures near the grain boundaries and precipitates. (a) CW34. (b) HT450-5.

The second effect of annealing is shown in the STEM images of the CW34 HT450-5 and HT450-10 wires depicted in Figs. 6a, 6b and 6c, respectively. The main RE elements in the WE43 alloy (Nd and Y) are homogeneously distributed within the CW34 wire, and there is no evidence of segregation at either grain boundaries or dislocations (Fig. 6a). Short annealing during 5s at 450ºC after cold drawing led to an important reduction in the dislocation density (as indicated above) and to the segregation of Nd and, to a minor extent, Y at the grain boundaries (Fig. 6b). Further annealing up to 10s at 450 ºC, led to the nucleation of small precipitates rich in Nd and Y near the grain boundaries (Fig. 6c).

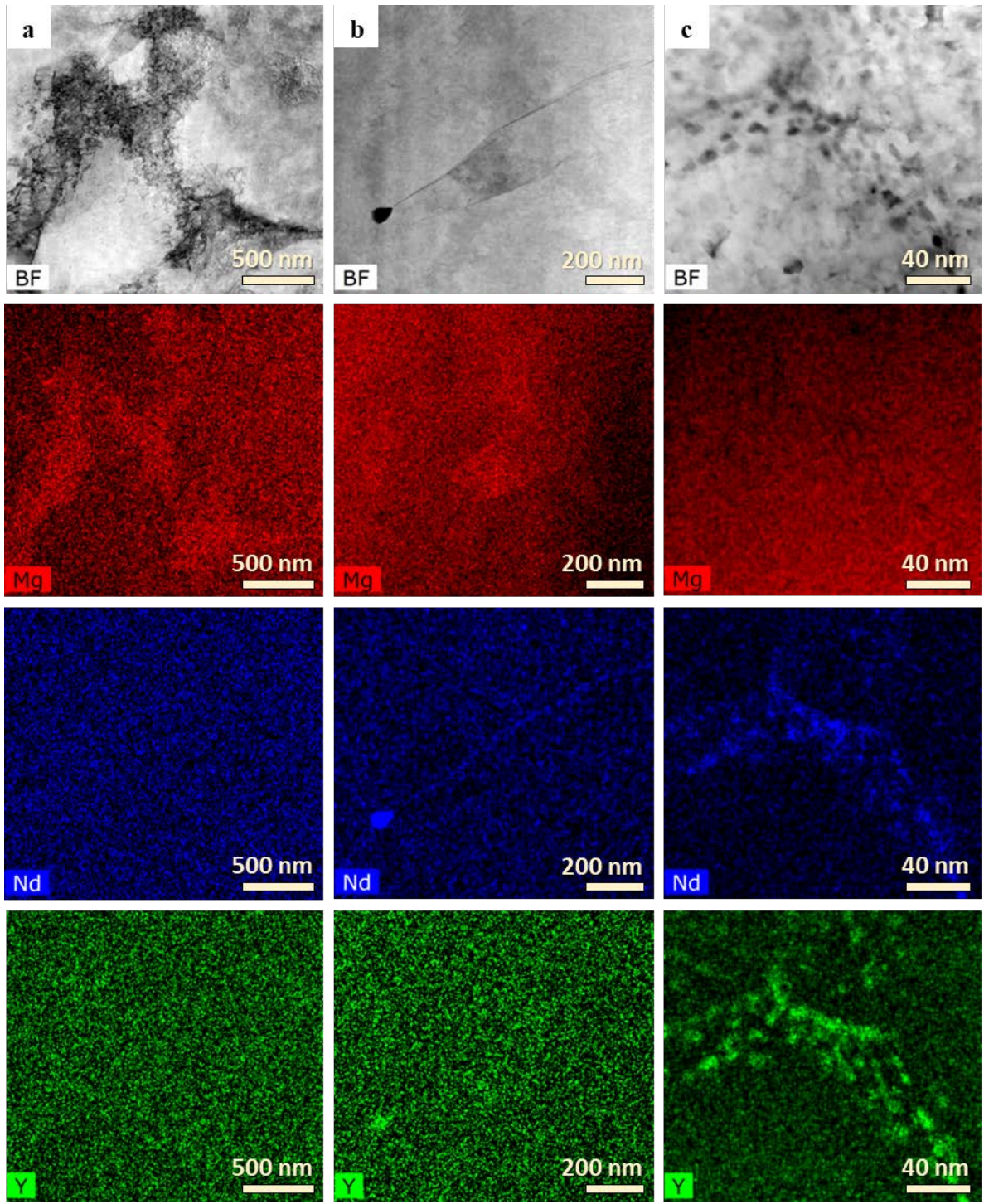

**Fig 6.** STEM images and EDS maps showing the distribution of Mg, Nd, and Y. (a) CW34. (b) HT450-5 and (c) HT-450-10.

Representative engineering tensile stress-strain curves of the WE43 Mg wires subjected to different processing conditions are depicted in Fig. 7a. The average values and standard deviation of the 0.2% offset yield strength, tensile strength and strain to failure are summarized in Fig. 7b. The stress was obtained from the load divided by the initial cross-section of the wire while strains was calculated from the crosshead displacement of the mechanical testing machine divided by the initial gage length. The wires that were not annealed after cold drawing (CW34 and CW13) showed high strength and very poor ductility (< 1%). This brittle behavior agrees with the microstructure of the wires, which showed small grain size (≈ 1 μm) and large dislocation densities within the grains associated with the cold drawing process (Fig. 5). Very large external stress has to be applied to promote dislocation motion against the back stresses created by the dislocation pileups at the grain boundaries, leading to brittle failure of the wires. It should be noted that very similar results were obtained for both 13% and 34% of cold work, indicating that 13% cold work is sufficient to promote strain hardening of the wires.

Annealing of the CW13 Mg wires at either 400ºC or 450ºC during 5 s or 10 s however led to significant changes in the stress-strain behaviour. The yield strength dropped slightly, and this reduction increased with both temperature and annealing time, while the ductility increased up to 7 % after annealing at 400ºC and up to ≈ 10% after annealing at 450ºC. Annealing did not lead to significant changes in texture or grain size but dramatically reduced the dislocation density within the grains (Figs. 3 and 5) and thus allowed the activation of plastic slip at lower stresses. Further plastic deformation led to strain hardening and improved the strength and ductility of the wires. Among the different annealing treatments, 450ºC during 5 s led to the optimum combination of strength and ductility according to the data in Figs. 7a and 7b.

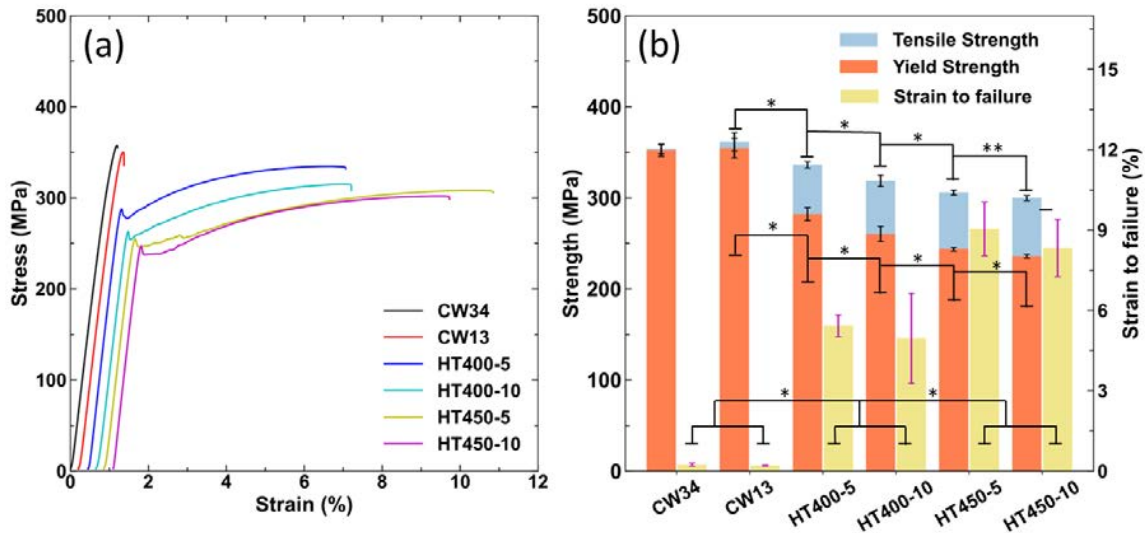

**Fig. 7.** (a) Representative engineering tensile stress-strain curves of WE43 Mg wires undergoing different processing conditions. The different curves are displayed with an offset ≈ 0.2% for the sake of clarity. (b) Summary of tensile properties of the Mg wires with different processing conditions. Here (*) and (**) represents $p < 0.05$ and $0.05 < p < 0.10$, respectively obtained by one-way ANOVA.

### 3.2. Effect of surface-modification by PEO

PEO surface modification is an electrochemical conversion process based on principles of plasma electrolysis, also known as micro-arc oxidation (MAO) [41][36]. During PEO, the application of a high electrical potential results in strong, but short-lived plasma discharges associated high a significant local increase in temperature, leading to a uniform effusion of molten Mg through discharge channels at the surface of the substrate. The molten Mg rapidly reacts during discharge with the chemical compounds of the electrolyte and is immediately quenched by the cold electrolyte bath. An oxide layer builds up within short periods of time because of those plasma ignitions scanning the surface regions exposed to the electrolyte. However, as the oxide layer on the surface grows, the electrical resistance between the anode and cathode increases up to a critical value in voltage, which finally needs to be surpassed by the electrical source or automatically limits film growth and terminates the process. From a chemical standpoint, the oxide layer induced by PEO surface modification comprises contributions from the Mg substrate, electrolyte species which react or are incorporated as well as in some cases from alloying elements. Conclusively, PEO of Mg alloys in the presence of a phosphate-based electrolyte at least leads to oxide rich layers which contain both MgO and $Mg_3(PO_4)_2$ [33][35].

The continuous surface-modification by PEO was only performed on the HT450-5 wires to study the basic principles of continuous processing as well as to analyze the most influential parameters. From a qualitative viewpoint, many plasma sparks appeared around the wire as it entered the electrolyte bath during PEO, but the density of sparks decreased and became negligible when the wire exited the bath, as schematically shown in Fig. 2. To understand the formation of the surface modification layer, the continuous process was stopped and the cross-sections and lateral surfaces of wire segments from different locations inside the bath were analyzed by optical microscopy and SEM. Images of the wire surface are depicted in Figs. 8a, 8b and 8c after 3, 9 and 20 s in the bath, respectively, shown at higher magnification in Figs. 8d, 8e and 8f. The apparent color of the wire turned from silver to off-white with a smooth appearance which is associated with the formation of the porous surface modification and it's ceramic nature [35]. The evolution of the surface modification layer with time is shown in Figs. 8g to 8i. These figures indicate that the interface in the facilitated process grows rapidly and is linked to the high density of sparks that are appear when the wire enters the bath. The morphology of the oxide layer, as displayed in Figs. 8h and 8i, is very similar and the only difference is the thickness whose evolution with time in the electrolyte has been plotted in Fig. 9 for PEO processes carried out at a frequency of 250 Hz (PEO 250) and 500 Hz (PEO 500). The layer initially grows rapidly in the first 5 s but apparently, the thickness increases only slowly after 10s. The frequency has very limited influence on the oxide layer thickness after 20 s, leading to average thicknesses of the oxide layers of 8.3 ± 1.3 µm and 7.8 ± 1.3 µm for PEO250 and PEO500, respectively. It was assumed that this layer thickness was appropriate to passivate the Mg wire and further analyses of the morphological features were carried out on wires that spent 20 s in the electrolyte bath during the continuous PEO.

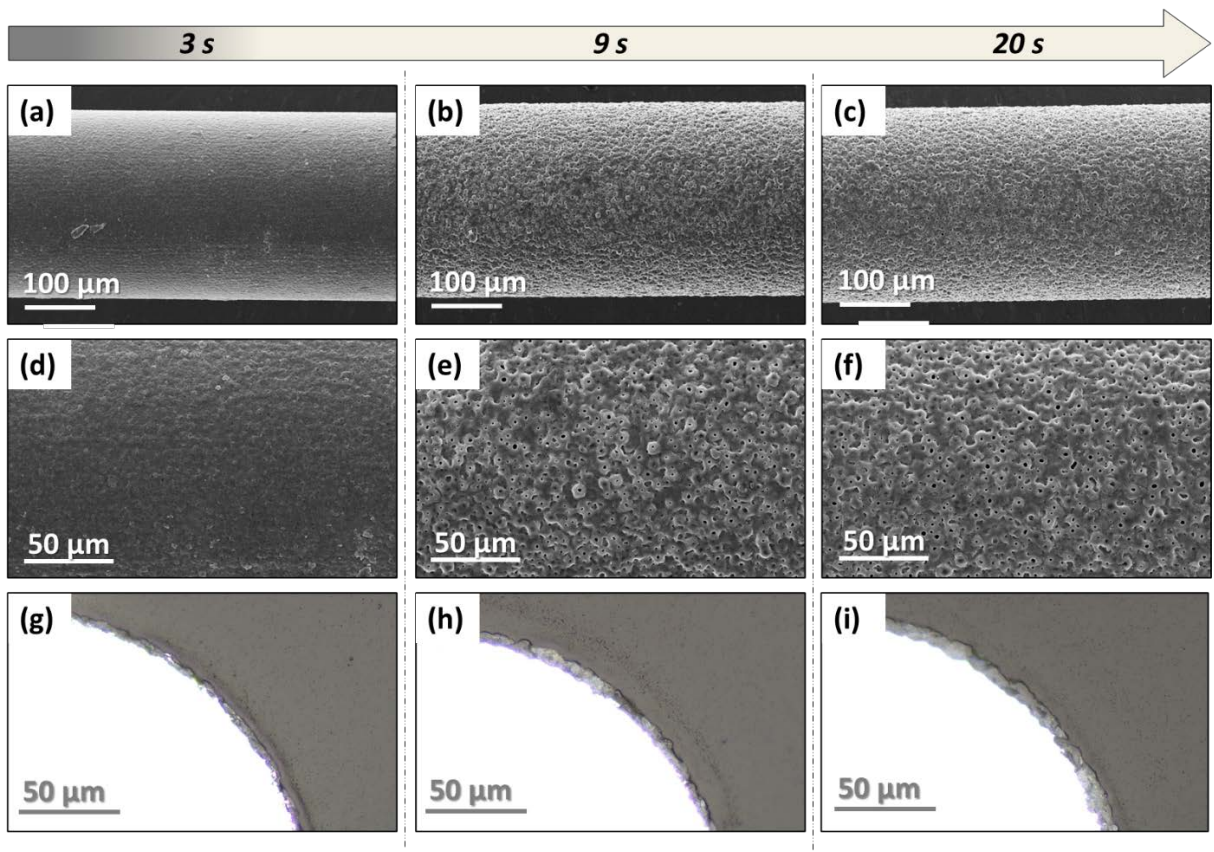

**Fig. 8.** (a-c) Secondary electron SEM image of the oxide layer formed by PEO on the wire surface as a function of time inside the electrolyte bath. (d-e) *Idem* at higher magnification. (g-h) Backscattered electron SEM images of transverse cross-section of the wire as a function of time inside the electrolyte bath. The frequency of the current during PEO was 500 Hz. The arrow on top of the figures shows the evolution of the actual color of the Mg wire with time.

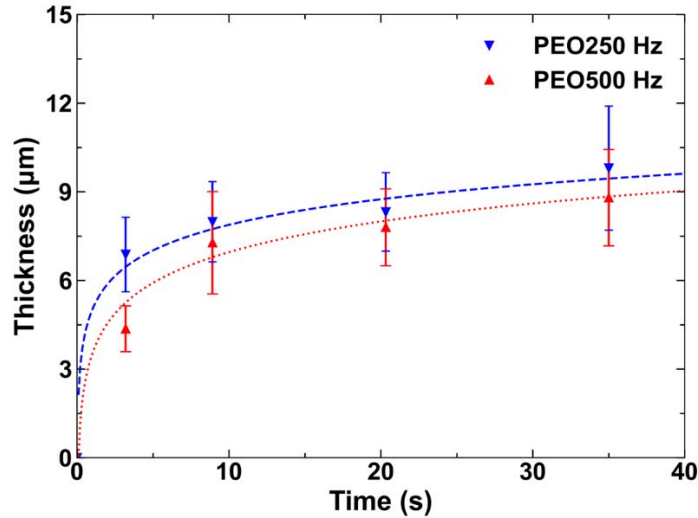

**Fig. 9.** Evolution of oxide layer thickness as a function of the time inside the electrolyte bath carried out at 250 Hz (PEO250) and 500 Hz (PEO500).

The surface morphology of the continuous oxide layer obtained by 20 s immersion in the electrolyte at a frequency of 250 Hz and 500 Hz is shown at high magnification in the secondary electron SEM images in Figs. 10a and 10b, respectively. The structure of the modification layer perpendicular to the wire axis can be observed in the backscattered electron images of the transverse cross-section depicted in Figs. 10c and 10d. There is a sharp interface between the Mg wire (characterized by the presence of many globular β' precipitates that appear white in the backscattered electron images) and the oxide layer. The oxide layer grown at 250 Hz shows large pores in the middle of the oxide layer, while PEO processing at 500 Hz led to a more compact interface with reduced internal porosity. The analysis of the longitudinal sections of the wires observed by light microscopy (Figs. 10e and 10f) demonstrates that the thickness of the oxide layer as well as the microstructure are constant along the wire length, indicating that the facilitated continuous PEO process is robust and can be applied on continuous wires for serial production.

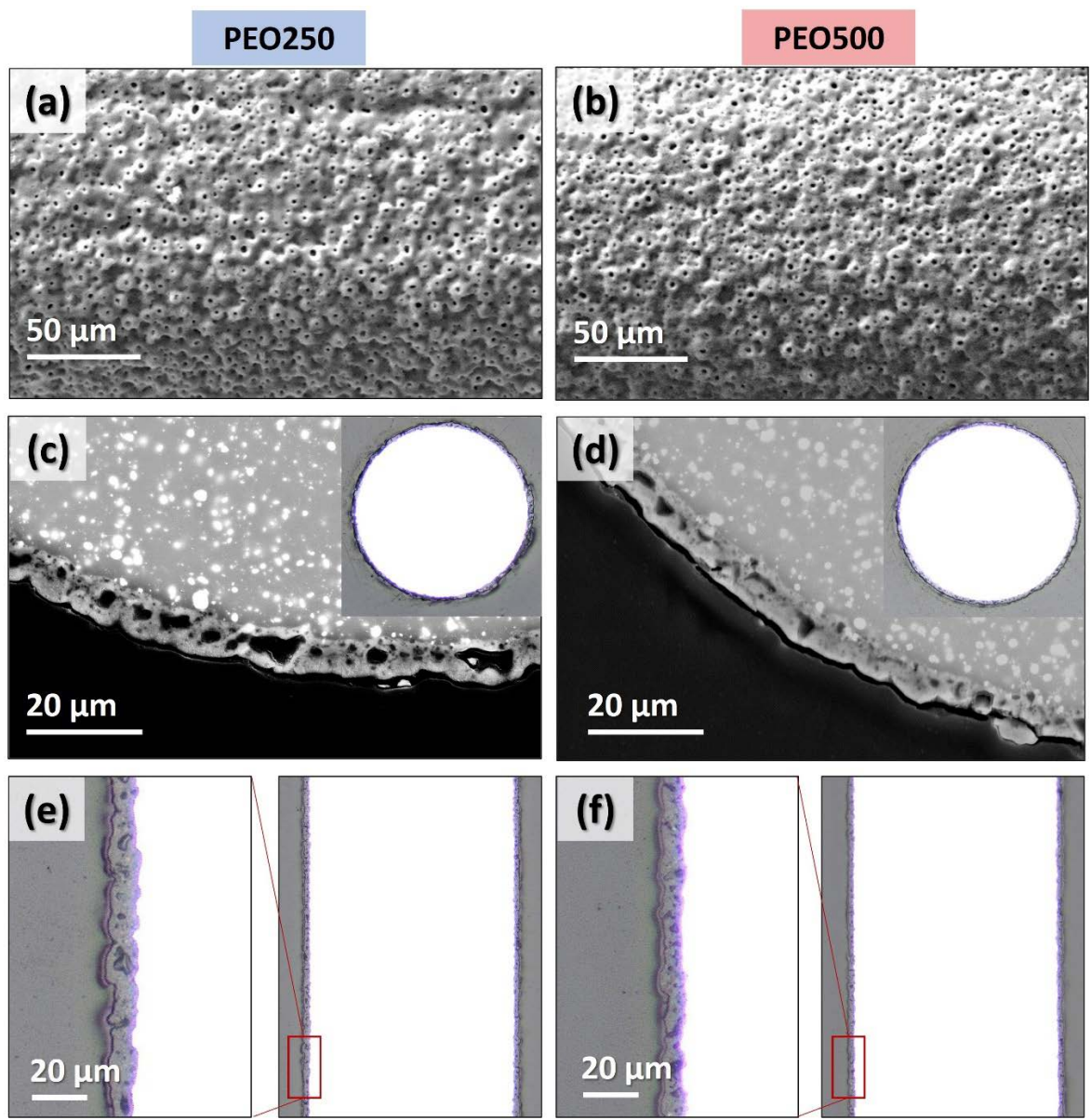

**Fig. 10.** (a) Secondary electron SEM image of the oxide layers formed at 250 Hz. (b) *Idem* at 500 Hz. (c) Backscattered electron SEM images of transverse cross-section of the wire after PEO at 250 Hz. (d) *Idem* at 500 Hz. (e) Light microscopical image of the longitudinal section of wire after PEO at 250 Hz. (f) *Idem* at 500 Hz. The average thicknesses of the oxide layers were 8.3 ± 1.3 µm and 7.8 ± 1.3 for PEO250 and PEO500, respectively.

The chemical composition of the oxide layer can be found in the element composition maps obtained by EDX in the cross-section of the PEO500 wire in Fig. 11. They confirm the presence of Mg, P and O, in agreement with previous reports [33][35]. The most likely chemical compounds in the oxide rich interface layer are MgO and $Mg_3(PO_4)_2$, which appear because of the following reactions:

$$Mg \rightarrow Mg^{2+} + 2e^- \quad (1)$$

$$4OH^- \rightarrow 2H_2O + O_2 + 4e^- \quad (2)$$

$$Mg^{2+} + 2OH^- \rightarrow Mg(OH)_2 \quad (3)$$

$$Mg(OH)_2 \rightarrow MgO + H_2O \quad (4)$$

$$3Mg^{2+} + 2PO_4^{3-} \rightarrow Mg_3(PO_4)_2 \quad (5)$$

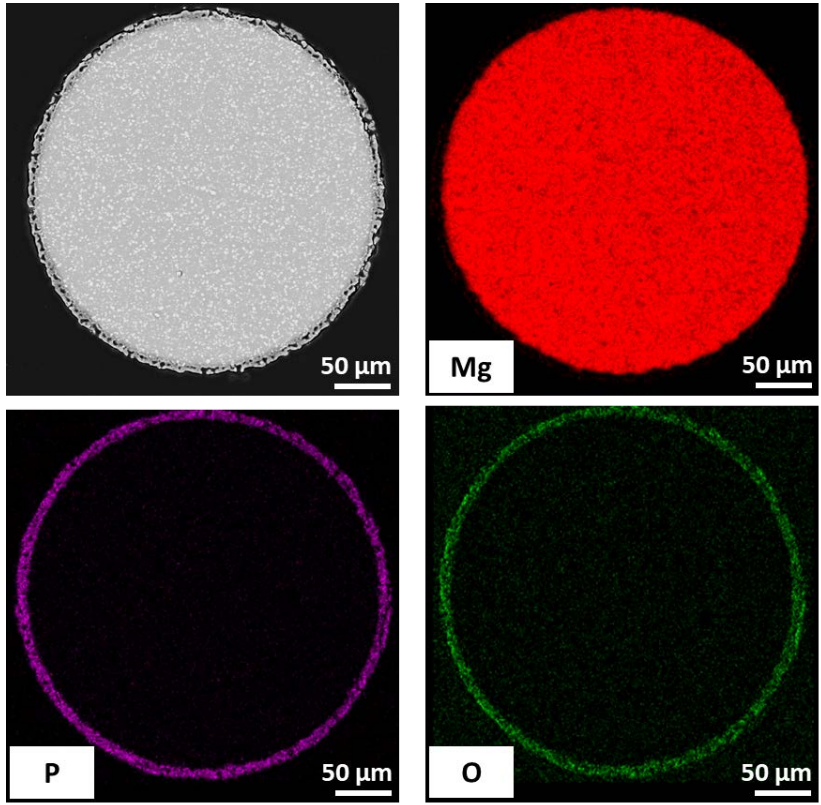

**Fig. 11**. Back-scattered electron SEM image of the cross-section of the PEO500 (top left) and elemental composition mapping obtained by EDX showing the presence of Mg, P and O in the oxide layer.

From a production perspective, cold drawing of metallic wires can lead to the formation of unidirectional grooves and sub-micron cracks at the surface due to deterioration of the die by wear, poor lubrication, etc. These defects do not significantly alter the tensile properties of wire if they are below a certain threshold because plastic deformation alleviates the stress concentrations. However, they are prone to undergo and promote pitting corrosion [21] and, in addition, contaminations can get into such small cavities making it difficult to eliminate by cleaning, thus potentially compromising biocompatibility. Examples of these longitudinal defects can be found in Fig. 12a. After the PEO process, the depth of the largest defects has been greatly reduced while the minor unidirectional drawing lines are entirely removed from the surface. This is because the oxide layer grows outwards and inwards simultaneously, at least at the beginning of the PEO process [42]. Our observations indicate that the total inwards growth was about 1-2 μm and was independent of the frequencies employed for PEO process, at least within the selected time span.

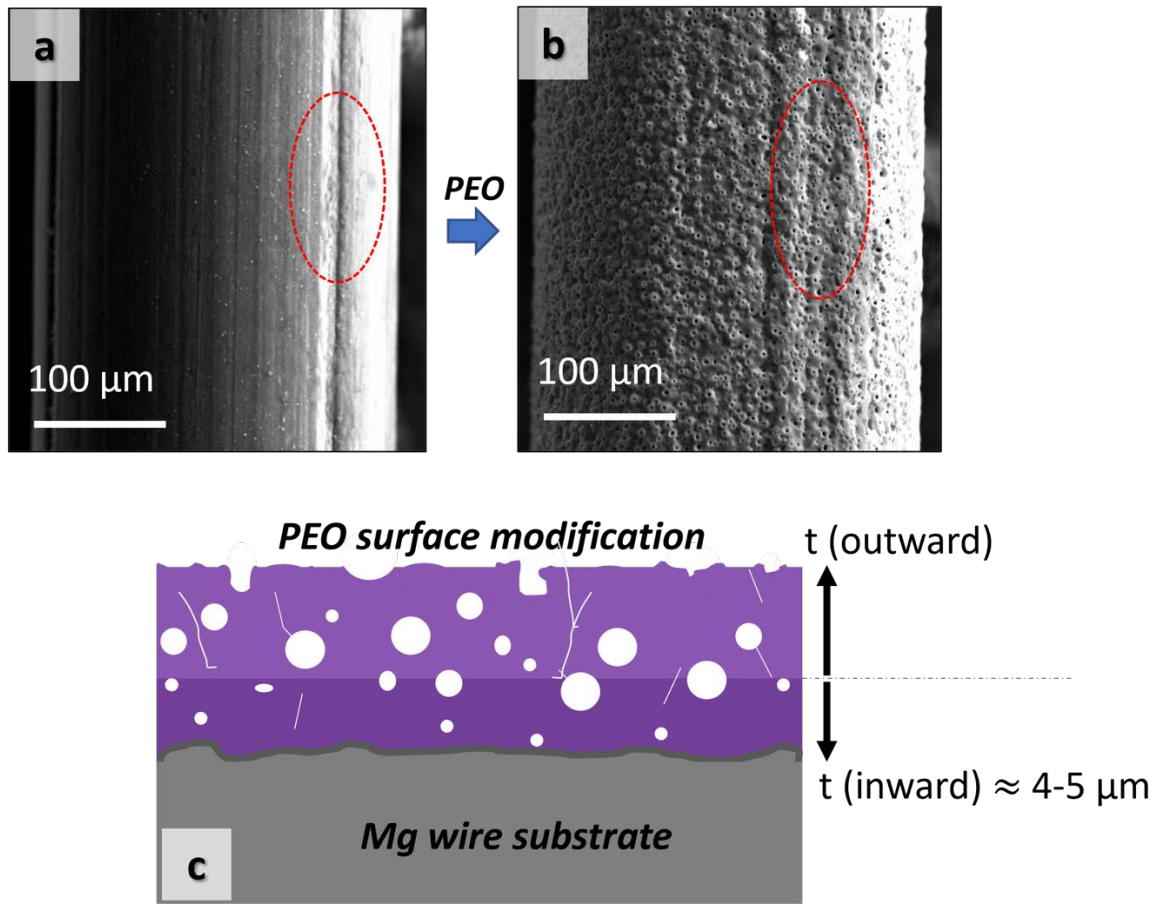

**Fig. 12**. (a) Secondary electron SEM image of Mg wire with longitudinal grooves created during cold drawing (b) Partial healing of the longitudinal grooves because of PEO. (c) Schematic depiction of the defect healing mechanism because of the inward penetration of the oxide layer.

Representative tensile stress-strain curves of the Mg wires before (HT450-5) and after (PEO250 and PEO500) surface modification are plotted in Fig. 13a. Surface-modification leads to a slight reduction in the strength of the wires for two reasons. Firstly, the PEO process consumes a distinct portion of Mg at the surface of the wire of 4-5 μm in thickness (Fig. 12), which is replaced by a porous oxide layer with higher strength, but limited ductility with regards to mechanical properties. Secondly, PEO is accompanied by an increase in the temperature of the wire, which leads to further annealing, potentially reducing the dislocation density and the yield strength. Nevertheless, the average results and standard deviation of yield strength, tensile strength, and ductility before and after PEO in Fig. 2b show that the effect of PEO is limited and does not significantly impair the mechanical performance of the tested Mg wires.

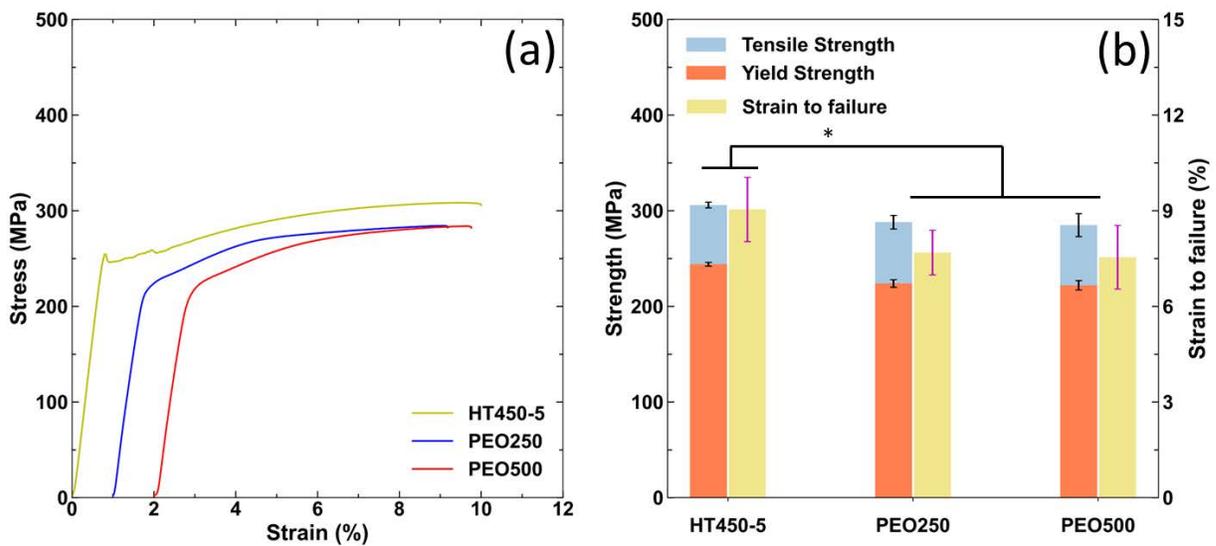

**Fig. 13.** Tensile stress-strain curves of Mg wires before (HT450-5) and after (PEO250 and PE0500) surface modification. The different curves are depicted with an off set of ≈ 1 % for the sake of clarity. (b) Summary of tensile properties of the Mg wires before (HT450-5) and after (PEO250 and PEO500) surface modification. Here, (*) represents $p < 0.05$ obtained by one-way ANOVA.

4. Discussion

In our work, Mg wires were successfully cold drawn from 1.50 mm to 0.30 mm in diameter in 28 steps and continuously subjected to different annealing heat treatments. The resulting cold drawn wires presented small grain size (~ 1 µm) and a marked texture with the basal planes parallel to the drawing axis. These wires were subjected to annealing treatments (400ºC/450ºC for 5/10 s) for stress relief, being mild enough to avoid recrystallization which would reduce the mechanical properties significantly. In fact, we found that the annealing treatments had limited effect on the grain size or the texture but reduced the dislocation density and led to the nucleation of precipitates which were rich in Nd and Y. While lower dislocation densities should be beneficial to reduce the corrosion rate, the presence of precipitates is expected to increase the corrosion rate in simulated body fluid because of the galvanic effect of the precipitates [31][38]. However, it should be noted that metallurgical processes are unlikely to reduce the corrosion rates of thin Mg wires [22] below the thresholds needed for biomedical applications.

Cold drawn wires that were not annealed after cold drawing showed a brittle behavior with high strength and very low ductility, regardless of the amount of cold work. This behavior was associated with the small grain size, strong texture and high dislocation density induced by the cold drawing process. Gentle annealing at 400ºC-450ºC for 5 or 10 s after cold drawing did not significantly modify the texture or the grain size but reduced the dislocation density within the grains [43]. As a result, strain hardening by accumulation of dislocations during plastic deformation was active and the annealed wires showed substantial ductility (in the range 5% to 10%) and high strength (> 300 MPa).

To overcome this limitation, a continuous PEO process was developed to modify the surface of the Mg WE43MEO wires in a continuous way to prove applicability to upscaling and serial production. The continuous PEO process allowed production of Mg wires for biomedical devices overcoming the limitation of the conventional PEO process which can only handle discrete elements (or small segments in case of wires [44]). In the conventional PEO process, the element is always in contact with the anode and the quality of surface modification is compromised at the contact point while the wire passes freely through the electrolyte during continuous PEO leading to a uniform surface modification along the wire. Moreover, the myriad of studies on conventional PEO of Mg alloys suggest that the processing time is generally in the order of minutes. For instance, 15 min, 10 min and 2.5~30 min were used by Chu et al [37], Monica et al [45], and Lv et al [46], respectively. The continuous PEO process presented in this investigation was able to produce a continuous and homogeneous PEO oxide layer in a few seconds, a crucial development for large-scale production of surface-modified Mg wires.

The coiled wire was continuously fed through an electrolyte bath only for few seconds in the continuous PEO process (best results were obtained for 20 s) which lead to the continuous formation of a porous yet resistant oxide layer of 8.3 ± 1.3 μm (PEO250) and 7.8 ± 1.3 (PEO500) in thickness which is in agreement with the report of Lee et al, where frequency had minor influence on the PEO oxide layer thickness [47]. EDX maps indicated that the layer was most likely made up of MgO and $Mg_3(PO_4)_2$. For PEO processes, the composition, morphology and porosity of the surface modification layer are important to control the corrosion resistance and can be tailored by varying the electrolyte composition, electrical parameters (voltage, duty cycle, frequency, current, etc.) and process time [33]. Because the current frequency is known to be a critical parameter to optimize PEO surface modifications [48], the effect of this factor was further investigated. It was found that increasing frequency reduced the pore size and increased film density (Fig. 10) while the growth rate of the oxide layer was slightly reduced (Fig. 8). These changes have been attributed to a reduction in the dimensions of the discharge channels (and to an increase in their number) when the frequency increases [33]. Hwang et al. [49] also reported a similar trend on oxide layers induced by PEO where the increase in frequency resulted in improved corrosion resistance. As a result, the corrosion resistance of the wires whose surface was modified by PEO showed a strong tendency to increase with frequency [48][50]. Similar results are reported in the second part of our work [35], in which higher frequencies and enhanced oxide layer generation showed important improvements in the strength after in vitro degradation and in the cell-wire interactions [51].

## 5. Conclusion

A novel strategy to manufacture Mg wires of 0.30 mm in diameter by cold drawing followed by continuous surface treatment by plasma electrolytic oxidation is presented. The novel strategy is easily scalable to produce large quantities of surface enhanced Mg wires for biomedical applications with significantly improved corrosion resistance and, thus, improved mechanical properties and cytocompatibility due to the presence of a continuous PEO oxide layer.

The cold draw process showed to be robust and wires up to lengths of 20 m were successfully manufactured. The cold drawn wires presented small grain size (around 1 μm) and strong basal texture, together with a large dislocation density and contained large, globular precipitates (in the range 0.5 to 1.5 μm) of $Mg_{12}YNd$ dispersed throughout the microstructure. Gentle annealing treatments at 400ºC/450ºC for 5/10 s were able to reduce the dislocation density (to improve the ductility) without inducing recrystallization (that may potentially impair mechanical properties). The annealing treatments led to the nucleation of Nd- and Y-rich nm-sized precipitates at the grain boundaries. It was found that the mechanical properties (in terms of strength and ductility) could be optimized by means of annealing treatments after cold-drawing at 450ºC during for 5s.

The cold drawing processes was combined with a novel continuous plasma electrolytic oxidation process which was able to create a homogeneous porous oxide layer made up of MgO and $Mg_3(PO_4)_2$ on the wire surface. The oxide layer thickness can be controlled by the time of the wire into the electrolytic bath, and thicknesses of ≈ 8 µm could be attained by PEO treatment for 20 s. In addition, the chemical composition, morphology and porosity of the oxide layer can potentially be tailored by changing the electrolyte composition and the electrical parameters. We were able to show that layers created at higher frequency presented lower porosity. Even though the PEO modification reduced the tensile properties of the wires slightly, sufficient strength of ≈ 300 MPa with a strain-to-failure ≈ 8% could ultimately be achieved for the first time using a continuous PEO process for biomedical Mg wires.

**Acknowledgments**

This investigation was supported by the European Union's Horizon 2020 research and innovation program under the European Training Network BioImplant (Development of improved bioresorbable materials for orthopedic and vascular implant applications), Marie Skłodowska-Curie grant agreement No 813869. Additional support from the Spanish Research Agency through the grant PID2021-124389OB-C21 also gratefully acknowledged.

**Conflict of interests**

Tillman, Mayer and Kopp are employed by Meotec GmbH. Presented magnesium wires are in development and commercially not available. The authors declare no conflict of interests.

**References**

[1] E. Scarcello, D. Lison, Are Fe-based stenting materials biocompatible? A critical review of in vitro and in vivo studies, J. Funct. Biomater. 11 (2020). https://doi.org/10.3390/jfb11010002.

[2] H. Guo, R.H. Cao, Y.F. Zheng, J. Bai, F. Xue, C.L. Chu, Diameter-dependent in vitro performance of biodegradable pure zinc wires for suture application, J. Mater. Sci. Technol. 35 (2019) 1662–1670. https://doi.org/10.1016/j.jmst.2019.03.006.

[3] M. Asgari, R. Hang, C. Wang, Z. Yu, Z. Li, Y. Xiao, Biodegradable metallicwires in dental and orthopedic applications: A review, 2018. https://doi.org/10.3390/met8040212.

[4] J. Xue, A.J. Griebel, Y. Zhang, C. Romany, B. Chen, J. Schaffer, T.P. Weihs, Influence of Thermal Processing on Resoloy Wire Microstructure and Properties, Adv. Eng. Mater. 23 (2021) 1–12. https://doi.org/10.1002/adem.202001278.

[5] C. Rendenbach, H. Fischer, A. Kopp, K. Schmidt-Bleek, H. Kreiker, S. Stumpp, M. Thiele, G. Duda, H. Hanken, B. Beck-Broichsitter, O. Jung, N. Kröger, R. Smeets, M. Heiland, Improved in vivo osseointegration and degradation behavior of PEO surface-modified WE43 magnesium plates and screws after 6 and 12 months, Mater. Sci. Eng. C. (2021) 112380. https://doi.org/10.1016/j.msec.2021.112380.

[6] J.M. Seitz, E. Wulf, P. Freytag, D. Bormann, F.W. Bach, The manufacture of resorbable suture material from magnesium, Adv. Eng. Mater. 12 (2010) 1099–1105. https://doi.org/10.1002/adem.201000191.

[7] S.K. Kang, R.K.J. Murphy, S.W. Hwang, S.M. Lee, D. V. Harburg, N.A. Krueger, J. Shin, P. Gamble, H. Cheng, S. Yu, Z. Liu, J.G. McCall, M. Stephen, H. Ying, J. Kim, G. Park, R.C. Webb, C.H. Lee, S. Chung, D.S. Wie, A.D. Gujar, B. Vemulapalli, A.H. Kim, K.M. Lee, J. Cheng, Y. Huang, S.H. Lee, P. V. Braun, W.Z. Ray, J.A. Rogers, Bioresorbable silicon electronic sensors for the brain, Nature. 530 (2016) 71–76. https://doi.org/10.1038/nature16492.

[8] R. Zan, W. Ji, S. Qiao, H. Wu, W. Wang, T. Ji, B. Yang, S. Zhang, C. Luo, Y. Song, J. Ni, X. Zhang, Biodegradable magnesium implants: a potential scaffold for bone tumor patients, Sci. China Mater. 64 (2021) 1007–1020. https://doi.org/10.1007/s40843-020-1509-2.

[9] W. Ali, M. Echeverry-Rendón, A. Kopp, C. González, J. LLorca, Strength, corrosion resistance and cellular response of interfaces in bioresorbable poly-lactic acid/Mg fiber composites for orthopedic applications, J. Mech. Behav. Biomed. Mater. 123 (2021) 104781. https://doi.org/10.1016/j.jmbbm.2021.104781.

[10] W. Ali, A. Mehboob, M.G. Han, S.H. Chang, Effect of fluoride coating on degradation behaviour of unidirectional Mg/PLA biodegradable composite for load-bearing bone implant application, Compos. Part A Appl. Sci. Manuf. 124 (2019) 105464. https://doi.org/10.1016/j.compositesa.2019.05.032.

[11] C. Redlich, A. Schauer, J. Scheibler, G. Poehle, P. Barthel, A. Maennel, V. Adams, T. Weissgaerber, A. Linke, P. Quadbeck, In vitro degradation behavior and biocompatibility of bioresorbable molybdenum, Metals (Basel). 11 (2021) 1–16. https://doi.org/10.3390/met11050761.

[12] S. Qiao, Y. Wang, R. Zan, H. Wu, Y. Sun, H. Peng, R. Zhang, Y. Song, J. Ni, S. Zhang, X. Zhang, Biodegradable Mg Implants Suppress the Growth of Ovarian Tumor, ACS Biomater. Sci. Eng. 6 (2020) 1755–1763. https://doi.org/10.1021/acsbiomaterials.9b01703.

[13] T.M. Hopkins, K.J. Little, J.J. Vennemeyer, J.L. Triozzi, M.K. Turgeon, A.M. Heilman, D. Minteer, K. Marra, D.B. Hom, S.K. Pixley, Short and long gap peripheral nerve repair with magnesium metal filaments, J. Biomed. Mater. Res. - Part A. 105 (2017) 3148–3158. https://doi.org/10.1002/jbm.a.36176.

[14] J. Xue, S. Singh, Y. Zhou, A. Perdomo-Pantoja, Y. Tian, N. Gupta, T.F. Witham, W.L. Grayson, T.P. Weihs, A biodegradable 3D woven magnesium-based scaffold for orthopedic implants, Biofabrication. 14 (2022). https://doi.org/10.1088/1758-5090/ac73b8.

[15] K. TESAŘ, K. BALÍK, Z. SUCHARDA, A. JÄGER, Direct extrusion of thin Mg wires for biomedical applications, Trans. Nonferrous Met. Soc. China (English Ed. 30 (2020) 373–381. https://doi.org/10.1016/S1003-6326(20)65219-0.

[16] M. Nienaber, S. Yi, K.U. Kainer, D. Letzig, J. Bohlen, On the Direct Extrusion of Magnesium Wires from Mg-Al-Zn Series Alloys, (2020).

[17] W. Roger N, Wire Technology: Process Engineering and Metallurgy, 2010. https://www.elsevier.com/books/wire-technology/wright/978-0-12-382092-1.

[18] A.L. Oppedal, H. El Kadiri, C.N. Tomé, S.C. Vogel, M.F. Horstemeyer, Anisotropy in hexagonal close-packed structures: Improvements to crystal plasticity approaches applied to magnesium alloy, Philos. Mag. 93 (2013) 4311–4330. https://doi.org/10.1080/14786435.2013.827802.

[19] J.Y. Wang, N. Li, R. Alizadeh, M.A. Monclús, Y.W. Cui, J.M. Molina-Aldareguía, J. LLorca, Effect of solute content and temperature on the deformation mechanisms and critical resolved shear stress in Mg-Al and Mg-Zn alloys, Acta Mater. 170 (2019) 155–165. https://doi.org/10.1016/j.actamat.2019.03.027.

[20] J. Wang, J.M. Molina-Aldareguía, J. LLorca, Effect of Al content on the critical resolved shear stress for twin nucleation and growth in Mg alloys, Acta Mater. 188 (2020) 215–227. https://doi.org/10.1016/j.actamat.2020.02.006.

[21] M. Zheng, G. Xu, D. Liu, Y. Zhao, B. Ning, M. Chen, Study on the Microstructure, Mechanical Properties and Corrosion Behavior of Mg-Zn-Ca Alloy Wire for Biomaterial Application, J. Mater. Eng. Perform. 27 (2018) 1837–1846. https://doi.org/10.1007/s11665-018-3278-x.

[22] A.J. Griebel, J.E. Schaffer, T.M. Hopkins, A. Alghalayini, T. Mkorombindo, K.O. Ojo, Z. Xu, K.J. Little, S.K. Pixley, An in vitro and in vivo characterization of fine WE43B magnesium wire with varied thermomechanical processing conditions, J. Biomed. Mater. Res. - Part B Appl. Biomater. 106 (2018) 1987–1997. https://doi.org/10.1002/jbm.b.34008.

[23] Y. TIAN, H. wei MIAO, J. lin NIU, H. HUANG, B. KANG, H. ZENG, W. jiang DING, G. yin YUAN, Effects of annealing on mechanical properties and degradation behavior of biodegradable JDBM magnesium alloy wires, Trans. Nonferrous Met. Soc. China (English Ed. 31 (2021) 2615–2625. https://doi.org/10.1016/S1003-6326(21)65680-7.

[24] H.F. Sun, H.Y. Chao, E. De Wang, Microstructure stability of cold drawn AZ31 magnesium alloy during annealing process, Trans. Nonferrous Met. Soc. China (English Ed. 21 (2011) s215–s221. https://doi.org/10.1016/S1003-6326(11)61581-1.

[25] N. Stanford, M.R. Barnett, The origin of "rare earth" texture development in extruded Mg-based alloys and its effect on tensile ductility, Mater. Sci. Eng. A. 496 (2008) 399–408. https://doi.org/10.1016/j.msea.2008.05.045.

[26] A.J. Griebel, J.E. Schaffer, Expanding Magnesium's Reach through Cold Drawing, in: 72nd Annu. Int. Magnes. Assoc. Conf., 2015.

[27] A.J. Griebel, J.E. Schaffer, Cold-drawn ZM21 and WE43 wires exhibit exceptional strength and ductility, Eur. Cells Mater. 28 (2014) 2.

[28] J. Bai, L. Yin, Y. Lu, Y. Gan, F. Xue, C. Chu, J. Yan, K. Yan, X. Wan, Z. Tang, Preparation, microstructure and degradation performance of biomedical magnesium alloy fine wires, Prog. Nat. Sci. Mater. Int. 24 (2014) 523–530. https://doi.org/10.1016/j.pnsc.2014.08.015.

[29] J. Xie, J. Zhang, Z. You, S. Liu, K. Guan, R. Wu, J. Wang, J. Feng, Towards developing Mg alloys with simultaneously improved strength and corrosion resistance via RE alloying, J. Magnes. Alloy. 9 (2021) 41–56. https://doi.org/10.1016/j.jma.2020.08.016.

[30] H. Azzeddine, A. Hanna, A. Dakhouche, L. Rabahi, N. Scharnagl, M. Dopita, F. Brisset, A.L. Helbert, T. Baudin, Impact of rare-earth elements on the corrosion performance of binary magnesium alloys, J. Alloys Compd. 829 (2020) 154569. https://doi.org/10.1016/j.jallcom.2020.154569.

[31] M. Li, F. Benn, T. Derra, N. Kröger, M. Zinser, R. Smeets, J.M. Molina-Aldareguia, A. Kopp, J. LLorca, Microstructure, mechanical properties, corrosion resistance and cytocompatibility of WE43 Mg alloy scaffolds fabricated by laser powder bed fusion for biomedical applications, Mater. Sci. Eng. C. 119 (2021) 111623. https://doi.org/10.1016/j.msec.2020.111623.

[32] P. Tian, X. Liu, Surface modification of biodegradable magnesium and its alloys for biomedical applications, Regen. Biomater. 2 (2015) 135–151.

https://doi.org/10.1093/rb/rbu013.

[33] G. Barati Darband, M. Aliofkhazraei, P. Hamghalam, N. Valizade, Plasma electrolytic oxidation of magnesium and its alloys: Mechanism, properties and applications, J. Magnes. Alloy. 5 (2017) 74–132. https://doi.org/10.1016/j.jma.2017.02.004.

[34] O. Jung, R. Smeets, P. Hartjen, R. Schnettler, F. Feyerabend, M. Klein, N. Wegner, F. Walther, D. Stangier, A. Henningsen, C. Rendenbach, M. Heiland, M. Barbeck, A. Kopp, Improved in vitro test procedure for full assessment of the cytocompatibility of degradable magnesium based on ISO 10993-5/-12, Int. J. Mol. Sci. 20 (2019). https://doi.org/10.3390/ijms20020255.

[35] A. Kopp, T. Derra, M. Müther, L. Jauer, J.H. Schleifenbaum, M. Voshage, O. Jung, R. Smeets, N. Kröger, Influence of design and postprocessing parameters on the degradation behavior and mechanical properties of additively manufactured magnesium scaffolds, Acta Biomater. 98 (2019) 23–35. https://doi.org/10.1016/j.actbio.2019.04.012.

[36] F. Simchen, M. Sieber, A. Kopp, T. Lampke, Introduction to plasma electrolytic oxidation-an overview of the process and applications, Coatings. 10 (2020). https://doi.org/10.3390/coatings10070628.

[37] C.L. Chu, X. Han, J. Bai, F. Xue, P.K. Chu, Fabrication and degradation behavior of micro-arc oxidized biomedical magnesium alloy wires, Surf. Coatings Technol. 213 (2012) 307–312. https://doi.org/10.1016/j.surfcoat.2012.10.078.

[38] W. Ali, M. Echeverry-Rendón, G. Dominguez, K. van Gaalen, A. Kopp, C. González, J. LLorca, Bioabsorbable WE43 Mg alloy wires modified by continuous plasma-electrolytic oxidation for implant applications. Part II: degradation and biological performance, (2022).

[39] N. Dodyim, K. Yoshida, T. Murata, Y. Kobayashi, Drawing of of magnesium magnesium fine wire and and medical medical application application of of drawn drawn wire wire, Procedia Manuf. 50 (2020) 271–275. https://doi.org/10.1016/j.promfg.2020.08.050.

[40] J.F. Nie, B.C. Muddle, Characterisation of strengthening precipitate phases in a Mg-Y-Nd alloy, Acta Mater. 48 (2000) 1691–1703. https://doi.org/10.1016/S1359-6454(00)00013-6.

[41] A.L. Yerokhin, X. Nie, A. Leyland, A. Matthews, S.J. Dowey, Plasma electrolysis for surface engineering, Surf. Coatings Technol. 122 (1999) 73–93. https://doi.org/10.1016/S0257-8972(99)00441-7.

[42] R.O. Hussein, X. Nie, D.O. Northwood, An investigation of ceramic coating growth mechanisms in plasma electrolytic oxidation (PEO) processing, Electrochim. Acta. 112 (2013) 111–119. https://doi.org/10.1016/j.electacta.2013.08.137.

[43] W. Ali, M. Li, L. Tillman, T. Mayer, C. González, J. LLorca,  and A. Kopp,

Bioabsorbable WE43 Mg alloy wires modified by continuous plasma-electrolytic oxidation for implant applications. Part I: Processing, microstructure and mechanical properties, (2022).

[44] X. Li, C. Shi, J. Bai, C. Guo, F. Xue, P.H. Lin, C.L. Chu, Degradation behaviors of surface modified magnesium alloy wires in different simulated physiological environments, Front. Mater. Sci. 8 (2014) 281–294. https://doi.org/10.1007/s11706-014-0257-5.

[45] M. Echeverry-Rendon, V. Duque, D. Quintero, S.M. Robledo, M.C. Harmsen, F. Echeverria, Improved corrosion resistance of commercially pure magnesium after its modification by plasma electrolytic oxidation with organic additives, J. Biomater. Appl. 33 (2018) 725–740. https://doi.org/10.1177/0885328218809911.

[46] G.H. Lv, H. Chen, L. Li, E.W. Niu, H. Pang, B. Zou, S.Z. Yang, Investigation of plasma electrolytic oxidation process on AZ91D magnesium alloy, Curr. Appl. Phys. 9 (2009) 126–130. https://doi.org/10.1016/j.cap.2007.12.007.

[47] S.J. Lee, L.H.T. Do, J.L. Lee, H.C. Peng, Process design of micro-arc oxidation coatings based on magnesium lithium alloy and their characteristics, Int. J. Electrochem. Sci. 12 (2017) 11256–11270. https://doi.org/10.20964/2017.12.64.

[48] Z. Shahri, S.R. Allahkaram, R. Soltani, H. Jafari, Optimization of plasma electrolyte oxidation process parameters for corrosion resistance of Mg alloy, J. Magnes. Alloy. (2018). https://doi.org/10.1016/j.jma.2018.10.001.

[49] I.J. Hwang, D.Y. Hwang, Y.G. Ko, D.H. Shin, Correlation between current frequency and electrochemical properties of Mg alloy coated by micro arc oxidation, Surf. Coatings Technol. 206 (2012) 3360–3365. https://doi.org/10.1016/j.surfcoat.2012.01.041.

[50] M.S. Butt, A. Maqbool, M. Saleem, M.A. Umer, F. Javaid, R.A. Malik, M.A. Hussain, Z. Rehman, Revealing the Effects of Microarc Oxidation on the Mechanical and Degradation Properties of Mg-Based Biodegradable Composites, ACS Omega. 5 (2020) 13694–13702. https://doi.org/10.1021/acsomega.0c00836.

[51] M. Echeverry-Rendon, F. Echeverria, M.C. Harmsen, Interaction of different cell types with magnesium modified by plasma electrolytic oxidation, Colloids Surfaces B Biointerfaces. 193 (2020) 111153. https://doi.org/10.1016/j.colsurfb.2020.111153.